\documentclass{elsart3}

\usepackage{graphicx}

\usepackage{amssymb}

\begin{document}

\begin{frontmatter}



\title{Sensitivity and pointing accuracy of the NEMO km$^3$ telescope}
\author[lns]{C. Distefano}\ead{carla.distefano@lns.infn.it}
\author{ for the NEMO Collaboration}

\address[lns]{Laboratori Nazionali del Sud, INFN, Catania, Italy}

\begin{abstract}
In this paper we present the results of Monte Carlo simulation studies on the
capability of the proposed NEMO km$^3$ telescope to detect high energy neutrinos.
We calculated the detector sensitivity to muon neutrinos coming from a generic 
point-like source.
We also simulated the lack of atmospheric muons in correspondence to the
Moon disk in order to determine the detector angular resolution and to
check the absolute pointing capability.
\end{abstract}

\end{frontmatter}

\section{Introduction}
\label{sec:introduction}

The NEMO Collaboration is
conducting an R\&D activity towards the construction
of a Mediterranean km$^3$ neutrino telescope \cite{nemo}.
In this work, we present a first study of the expected response of the
proposed NEMO km$^3$ detector to neutrinos from
point-like sources.
In section \ref{sec:sensitivity}, we describe the generation of the
atmospheric neutrinos and muons background.
We define the detector sensitivity and
the event selection criteria.
Eventually, we determine the detector sensitivity
to a muon neutrino flux from a generic point-like source.
In section \ref{sec:moon}, we present results on possible
Moon shadow detection. We then estimate the detector angular resolution
and we check the effects of errors in the detector absolute orientation in
the pointing accuracy.

The geometry of the km$^3$ telescope, simulated
in this work, is the NEMOdh-140 described in \cite{rosa}. 
The detector response is simulated using the ANTARES
simulation codes \cite{antares-codes}, taken into account the 
water optical parameters measured in the site of Capo Passero \cite{Capopassero}.

\section{Detector sensitivity to point-like sources}
\label{sec:sensitivity}


The detector sensitivity is here calculated following the
the Feldman \& Cousins approach \cite{Feldman}. In particular we
define the sensitivity as:
\begin{equation}
\left(\frac{d\varphi_\nu}{d\varepsilon_\nu}\right)_{90}=
\frac{\overline{\mu}_{90}(b)}{n_s}\left(\frac{d\varphi_\nu}{d\varepsilon_\nu}\right)_{0},
\end{equation}
where $\overline{\mu}_{90}(b)$ is the 90\% c.l. average upper limit for 
an expected background with {\it known} mean $b$
and $(d\varphi_\nu/d\varepsilon_\nu)_0$ is an arbitrary source spectrum 
predicting a mean signal $n_s$.

\subsection{Simulation of the atmospheric background}
\label{sec:bkg}

A sample of $7\cdot10^9$ atmospheric neutrinos have been generated
using the ANTARES generation code, in the energy range 
$10^2\div10^8$ GeV, with a spectral index X=2 and a $4\pi$ isotropic
angular distribution. The events are weighted to to the sum of the Bartol flux \cite{Agrawal96}
and of prompt neutrino {\tt rqpm} model \cite{Bugaev98} flux.
Atmospheric muons are generated at the detector, applying a
weighted generation technique. We simulated $5.5\cdot10^7$ events in the
energy range $10^2\div10^6$ GeV. The events are weighted to the 
Okada parameterization \cite{okada}, taking into account the depth of the
Capo Passero site ($D=3500$ m) and the flux variation inside the detector sensitive height.

\subsection{Criteria for atmospheric events rejection}
\label{sec:selection}

The used reconstruction algorithm is a robust track fitting procedure based on a maximization
likelihood method.
We use, as {\it goodness of fit} criterion, the variable:
\begin{equation}
\Lambda\equiv -\frac{\hbox{log}(\mathcal{L})}{N_{DOF}}+0.1(N_{comp}-1),
\end{equation}
where $\hbox{log}(\mathcal{L})/N_{DOF}$ is the log-likelihood 
per degree of freedom and $N_{comp}$ is the total number of 
$1^\circ$ compatible solutions \cite{antares-codes}. 
A quality cut $\Lambda>\Lambda_{cut}$ is 
applied together with other selection criteria:
the muon must be reconstructed as up-going;
the event must be reconstructed with a number of hits $N_{fit}>N_{fit}^{cut}$;
only events reconstructed in a circular sky region
centered in the source position and having a radius of
$r_{bin}$ are considered.
The optimal values of $\Lambda_{cut}$, $N_{fit}^{cut}$ and $r_{bin}$ are chosen to minimize
the detector sensitivity, taking
into account both atmospheric neutrino and muon background. 

\subsection{Sensitivity to a generic point-like source}

\begin{figure}[h]
\begin{center}
\vspace*{-0.8cm}
\includegraphics[width=6cm]{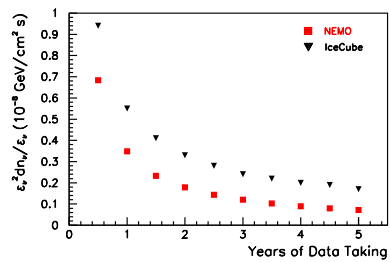}
\end{center}
\vspace*{-0.5cm}
\caption{Sensitivity to a neutrino spectrum with $\alpha=2$,
coming from a $\delta=-60^\circ$ declination point-like
source and comparison with the IceCube detector \cite{icecube}.}
\label{fig:point-sensi}
\end{figure}

We simulated muons induced by $\sim10^9$ neutrinos with energy range
$10^2\div10^8$ GeV and  X=1.
These events are weighted to the neutrino spectrum 
$(d\varphi_\nu/d\varepsilon_\nu)_0=10^{-7}\varepsilon_{\nu,GeV}^{-\alpha}$
(GeV$^{-1}$ cm$^{-2}$ s$^{-1}$). As a representative case,
the source position is chosen at a declination of $\delta=-60^\circ$.
In Table \ref{tab:sens-point-coin-0-180-9x9}, we report the expected sensitivity for different spectral index $\alpha$,
compared to the IceCube telescope \cite{icecube}. In Fig. \ref{fig:point-sensi} we plot the NEMO sensitivity
for $\alpha=2$ as a function of years of data taking, compared to the IceCube sensitivity 
obtained for a $1^\circ$ search bin \cite{icecube}.
The proposed NEMO detector shows a better 
sensitivity to muon neutrino fluxes, reached in a
smaller search bin.

\vspace*{0.1cm}
\begin{table}[h]
\begin{center} 
\begin{small} 
\begin{tabular}{|cccccc|c|} 
\hline 
\hline 
\multicolumn{6}{|c|}{NEMO} & IceCube       \\
\hline
$\alpha$ & $\Lambda_{cut}$ & $N_{fit}^{cut}$ & $r_{bin}$  &  $\bar{\mu}_{90}(b)$ & $\varepsilon_\nu^\alpha(d\varphi_\nu/d\varepsilon_\nu)_{90}$	& $\varepsilon_\nu^\alpha(d\varphi_\nu/d\varepsilon_\nu)_{90}$ \\ 
\hline 
1.0   &   7.6 &  26 & 0.4$^\circ$   &           2.43  &     $1.9\cdot10^{-15}$   & $2.4\cdot10^{-15}$ \\
1.5   &   7.3 &  -  & 0.5$^\circ$   &           2.45  &     $2.6\cdot10^{-12}$   & $4.5\cdot10^{-12}$ \\
2.0   &   7.3 &  -  & 0.5$^\circ$   &           2.45  &     $1.2\cdot10^{-9}$    & $2.4\cdot10^{-9}$ \\
2.5   &   7.3 &  -  & 0.6$^\circ$   &           2.48  &     $2.3\cdot10^{-7}$    & $3.8\cdot10^{-5}$ \\
\hline 
\hline 
\end{tabular} 
\end{small} 
\end{center} 
\caption{Sensitivity to a point-like neutrino source at $\delta=-60^\circ$, for different spectral indexes $\alpha$ and 3 years of
data taking, and comparison with the IceCube detector \cite{icecube}.  
The sensitivity spectrum $\varepsilon_\nu^\alpha(d\varphi_\nu/d\varepsilon_\nu)_{90}$ is expressed in GeV$^{\alpha-1}/$ cm$^2$ s.}   
\label{tab:sens-point-coin-0-180-9x9}
\end{table}

\section{Detection of the Moon shadow}
\label{sec:moon}

Since the Moon absorbs cosmic rays, we expect a lack of atmospheric muons from the direction of the
Moon disk (angular radius $R_{Moon}=0.26^\circ$). The detection of this muon deficit, commonly called {\it Moon shadow},
provides a measurement of the detector angular resolution; besides
detecting its position in the sky allows us to determine the absolute orientation of the
detector \cite{macro}. In this section, we present first results of our Monte Carlo simulations.

\subsection{Simulation of atmospheric muons}

Atmospheric muons are generated as described in section \ref{sec:bkg},
we implemented the calculation of the Moon position to simulate the muon lack.
We simulated $1.25\cdot10^8$ events in the energy range $10^2\div10^6$ GeV, 
restricting the generation  in a circular window around the Moon position
with a radius of 10$^\circ$. The events are then weighted to the 
Okada parameterization.

\subsection{Estimate of the detector angular resolution}

\begin{figure}[h]
\begin{center}
\vspace*{-0.5cm}
\includegraphics[width=6cm]{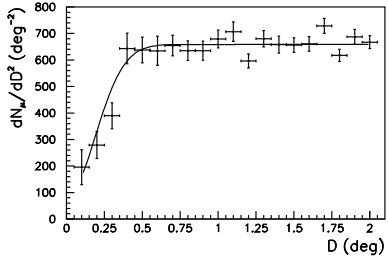}
\end{center} 
\vspace*{-0.5cm}
\caption{Selected muon event density versus the angular distance from the Moon centre. The line represents
the result of the fit in Eq.(\ref{eq:fit}), obtained for $k=(659\pm8)$ deg$^{-2}$ and $\sigma=0.19^\circ\pm0.2^\circ$.}
\label{fig:moon_plot}
\end{figure}

The number $N_\mu$ of muons reconstructed within a distance $D$ from the Moon centre 
is plotted in Fig. \ref{fig:moon_plot}. The distribution is then fitted using the function:
\begin{equation}
\frac{dN_\mu}{dD^2}=k\left( 1-\frac{R_{Moon}^2}{2\sigma^2}\exp\left(\displaystyle{\frac{D^2}{2\sigma^2}}\right)\right),
\label{eq:fit}
\end{equation}
obtaining from the fit an angular resolution $\sigma=0.19^\circ\pm0.2^\circ$.
During the fit procedure, we apply an event selection imposing $\Lambda_{cut}=-7.6$ and $N_{fit}^{min}=20$,
reaching, in this way, a detection significance $S_{\hbox{1yr}}\sim5.3$
in 1 year of data taking.
Assuming a required significance of 3, the minimum time needed to 
observe the shadow is therefore of the order of 100 days.

\subsection{Check of the detector pointing accuracy}

\begin{figure}[h]
\begin{center}
\includegraphics[width=2.8cm]{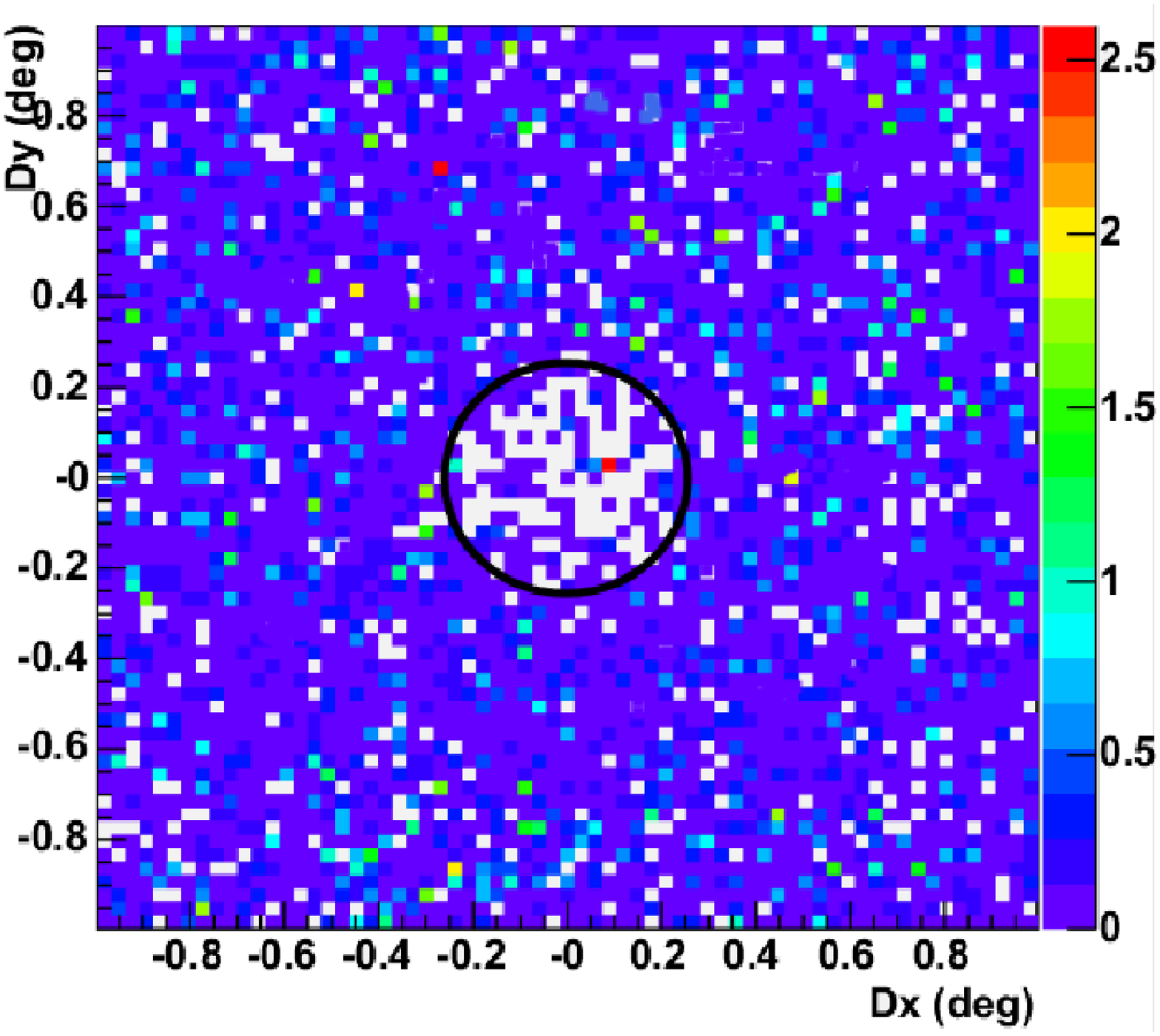}\includegraphics[width=2.8cm]{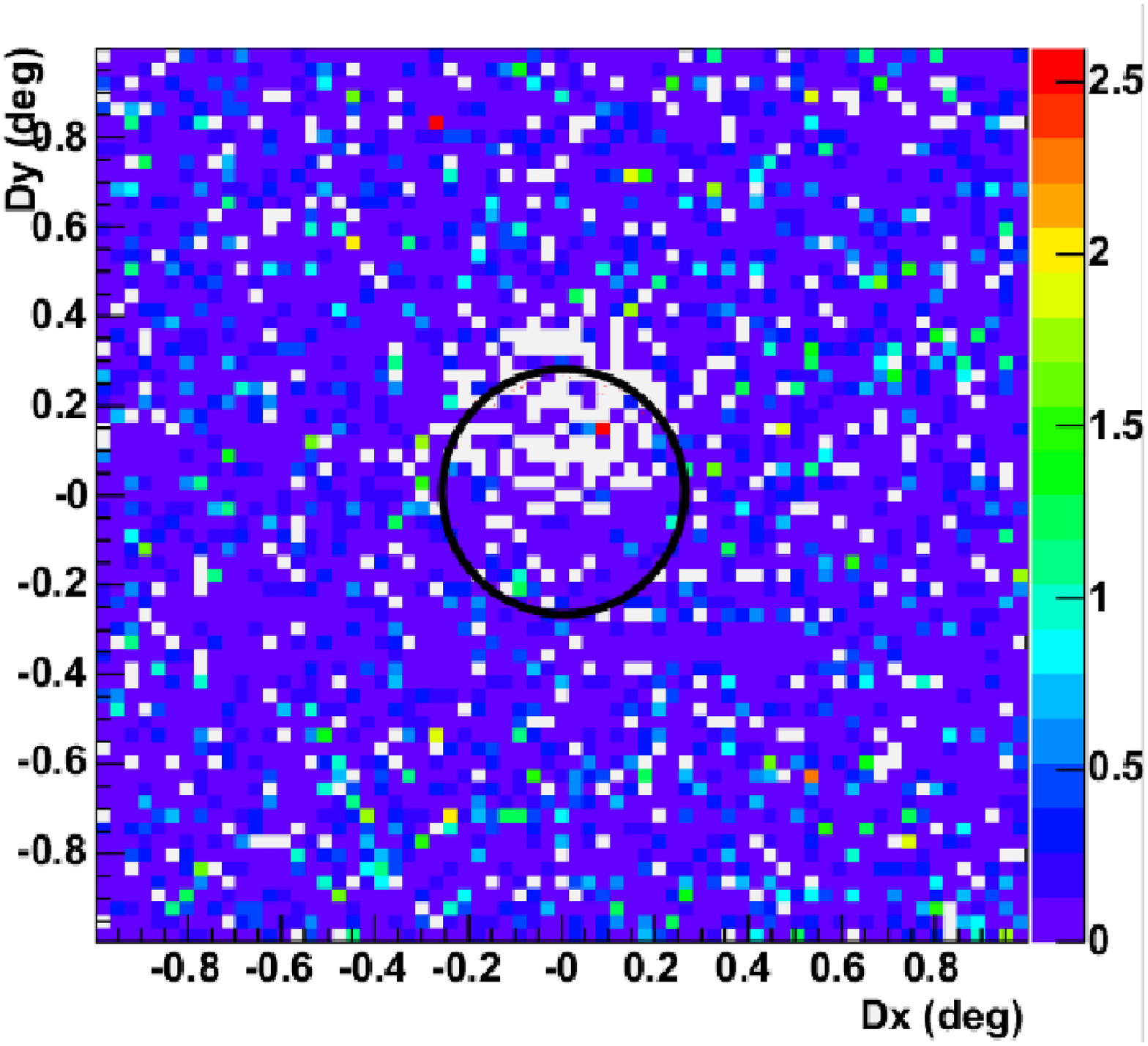}
\includegraphics[width=2.8cm]{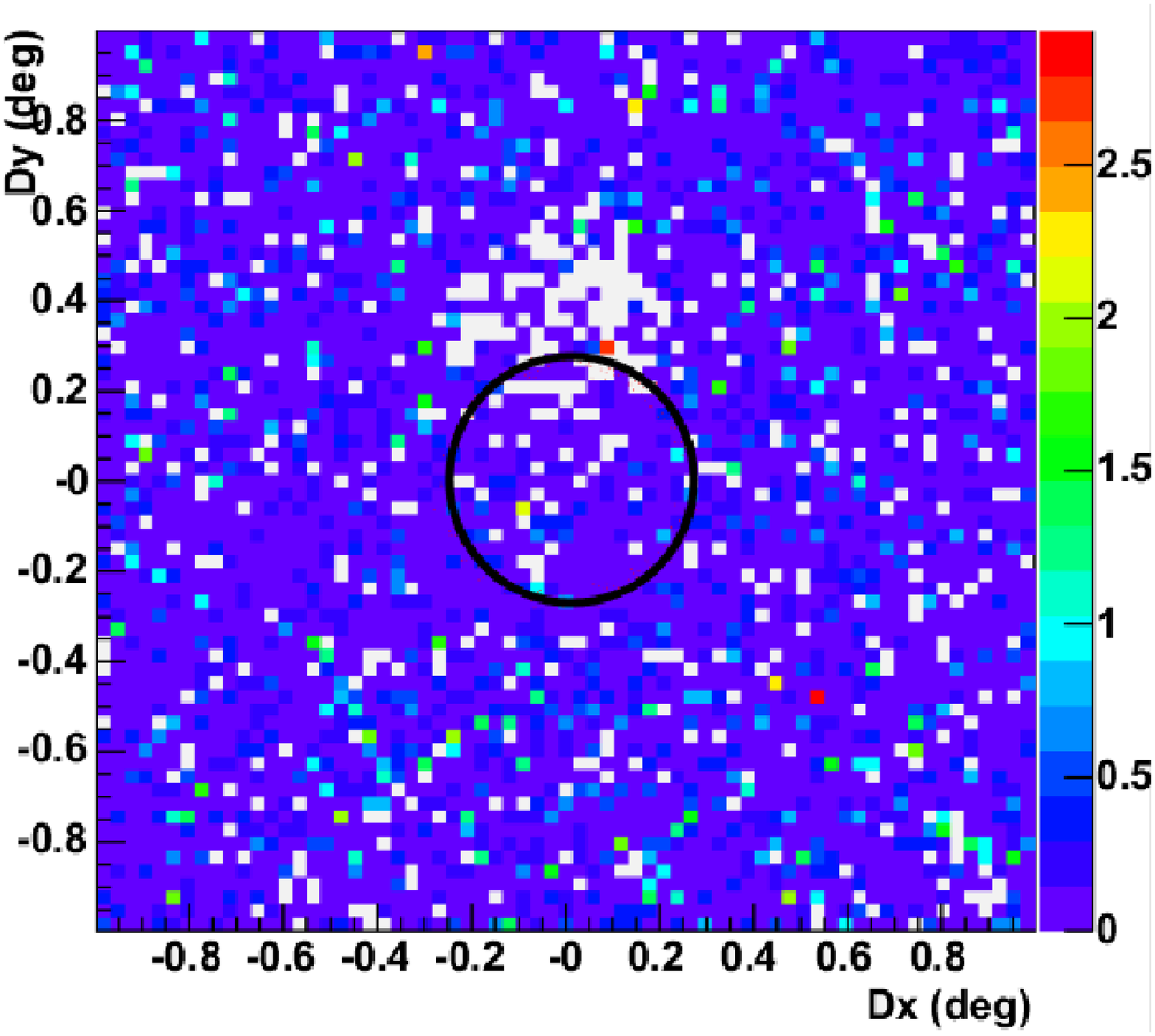}\includegraphics[width=2.8cm]{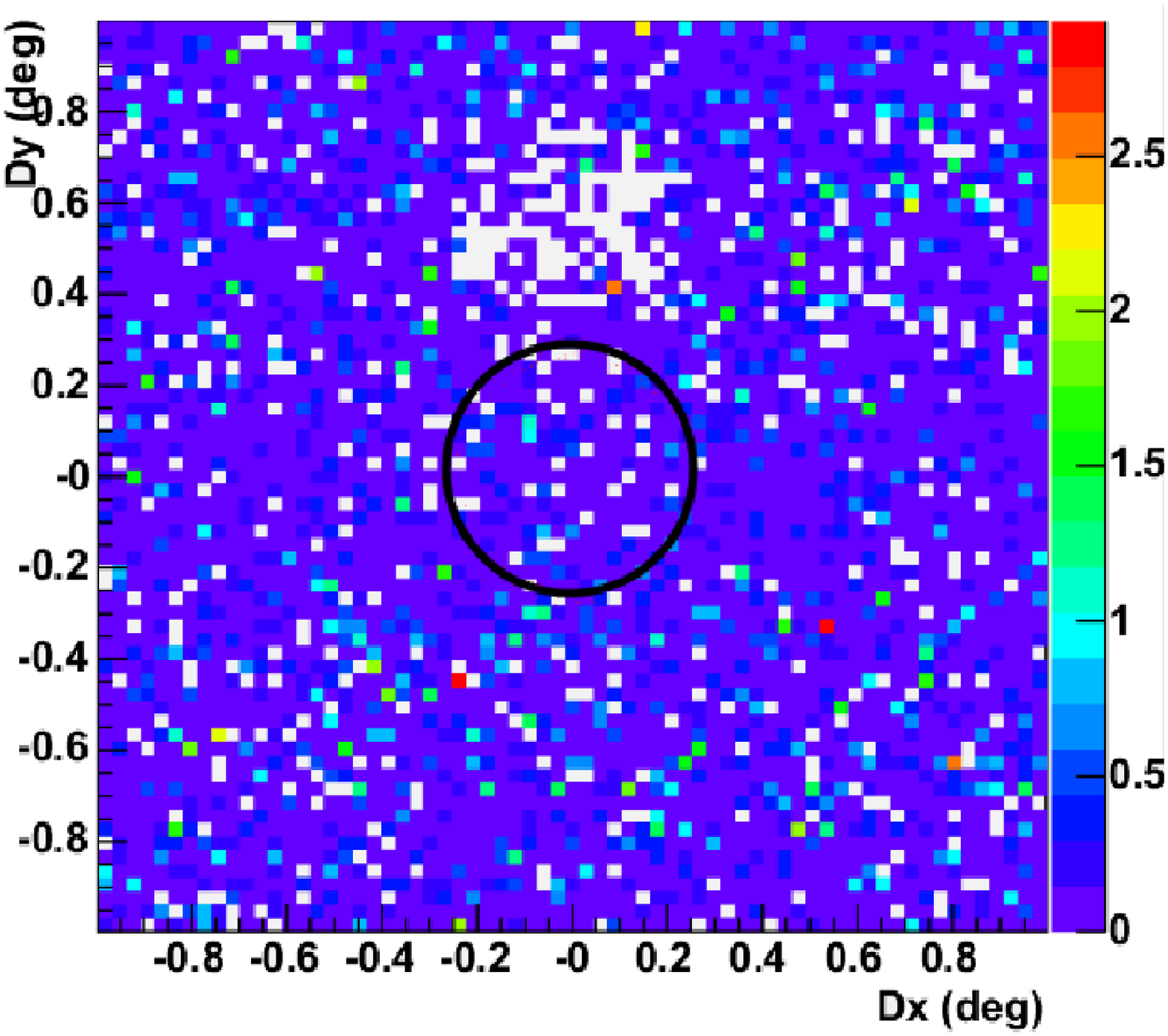}
\end{center}
\caption{Celestial map of selected events, referring to the Moon position, for $\Psi=0^\circ$ (top-left), $0.2^\circ$ (top-right),
$0.4^\circ$ (bottom-left) and $0.6^\circ$ (bottom-right). The black circle represents the Moon disk.}
\label{fig:moon_rotation}
\end{figure}

Up to now, we assumed to well know the detector absolute orientation. The effect of a possible
error in the absolute azimuthal orientation is here simulated, introducing a rotation $\Psi$
in the reconstructed tracks around the Z axis. In Fig. \ref{fig:moon_rotation}, we plot the selected
events in a 2D celestial map referring to the Moon position, for $\Psi=0^\circ,0.2^\circ,0.4^\circ$ and $0.6^\circ$.
For the expected accuracy $\Psi\leq0.2^\circ$, the shadow is still observable at the Moon position, 
while considering the pessimistic case $\Psi\geq0.2^\circ$, systematic errors may be corrected.
A detailed study together with the effects of errors in the absolute
zenithal orientation are still under analysis.

\begin{thebibliography}{00}
\bibitem{nemo} E. Migneco for the NEMO Collaboration, proc. in this workshop. 
The NEMO web site: nemoweb.lns.infn.it.
\end{thebibliography}
\end{document}